# Nonlinear Polarization Evolution of ultrashort pulses in Polarization Maintaining fibers

**JAN SZCZEPANEK,[1] TOMASZ M. KARDAŚ,[2] CZESŁAW RADZEWICZ,[1] AND YURIY STEPANENKO[2,*]**

[1]*Institute of Experimental Physics, Faculty of Physics, University of Warsaw, Pasteura 5, 02-093 Warsaw, Poland*
[2]*Institute of Physical Chemistry Polish Academy of Sciences, Kasprzaka 44/52, 01-224 Warsaw, Poland*
[1]*jan.szczepanek@fuw.edu.pl*
*[*]stepanenko@ichf.edu.pl*

**Abstract:** We examine properties of an ultrashort laser pulse propagating through an artificial Saturable Absorber based on Nonlinear Polarization Evolution device which has been realized with Polarization Maintaining fibers only (PM NPE). We study and compare in-line and Faraday Mirror geometries showing that the latter is immune to errors in the PM NPE construction. Experimental results for the transmission measurements of the PM NPE setup for different input linear polarization angles and various input pulse powers are presented. We show that PM NPE topology is of crucial importance for controlling the properties of the output pulse as it rules the contribution of cross-phase modulation to an overall nonlinear phase change. We also demonstrate an excellent agreement between the numerical model and experimental results.

## 1. Introduction

In this paper we demonstrate a theoretical model and experimental results of the intensity-dependent transmission process based on Nonlinear Polarization Evolution (NPE) [1] in Polarization Maintaining (PM) fiber structures. The process of the polarization state self-change due to the nonlinear effects was used for a number of purposes such as optical pulse reshaping experiments [2], optical discriminators [3,4], logic gates [5] as well as for demultiplexing of optical pulses [6]. In fiber lasers the mode-locking based on NPE process is an alternative to material Saturable Absorbers (SA). Passive mode-locking in all-fiber cavities has been demonstrated by means of different material SAs such as SESAM [7], graphene [8], or carbon nanotubes [9]. These SAs, however, tend to degrade during long-term operation, especially in fibers, under high intensity laser radiation [9]. One way to overcome potential damage of an SA is to use techniques based on nonlinear effects which can mimic saturable absorption process while assuring stable long-term performance. The original nonlinear SA technique in optical fibers called NPE is based on the filtration of the pulse polarization state affected by a self-action due to the Kerr nonlinearity. Recently, automated optimization of the NPE process based on bulk liquid crystals retarders placed in the cavity has been presented [10]. This approach resulted in hundreds or thousands of different mode-locking states in a single laser cavity. The enormous number of mode-locking states shows how critical the control of the initial polarization state is, when standard non-Polarization Maintaining optical fibers are used. For many years it was believed that NPE method is assigned uniquely to non-Polarization Maintaining [7,11] or partially PM fiber cavities in which NPE occurs in isotropic fiber sections [12]. The most prospective techniques are, however, these that can be used in all-fiber cavities build solely from PM fibers as they provide the high environmental stability of the laser [13]. Alternatives: the Nonlinear Optical Loop Mirror (NOLM) [14] or its variation Nonlinear Amplifying Loop Mirror (NALM) [15] are, however, to this day the most prominent artificial SAs used in environmentally stable all-PM-fibers cavities.

Environmentally stable and reliable lasers are required in both scientific and industrial applications. All-PM-fiber configurations with increased stability mode-locked with NOLM [13] and NALM [16] have been presented. Fermann et. al. presented NPE mode-locking using birefringent PM fiber between two polarizers, however, as in the standard NPE implementation, the mechanical adjustment of a proper polarization state inside the cavity was needed [17]. Additional increase of the environmental stability of fiber laser cavities mode-locked with various SAs has been achieved by using a Faraday Rotator (FR) and Faraday Mirror (FM) [18–20]. In this approach, an automatic compensation of unintentional polarization changes and Group Velocity Mismatch (GVM) between orthogonal polarizations is automatically assured. Vinegoni et al. presented the principle of PM NPE device using birefringent fibers where polarization state changes are induced by nonlinear effects only using quasi CW laser source [21]. Nielsen et al. showed the pulse generation in a linear all-PM-fiber cavity where NPE process took place in a very long (11 m) piece of birefringent fiber [22]. The cavity had a linear architecture employing an FM and produced 5.6 ps noise-like [23] pulses at 5.96 MHz repetition rate. Later, similar all-fiber configurations which have produced narrow bandwidth coherent picosecond pulses at low repetition rates from linear cavities were presented [24–26]. Recently, we have shown that PM NPE can be adapted for generation of 150 fs dissipative-soliton [27] ultra-short pulses in all-PM-fiber cavities [28]. However, the effective transmission of the SA in presented configuration was sensitive to errors in fiber segment lengths used for PM NPE process.

In this work we present in detail our new approach which is based on an appropriate sectioning of the birefringent PM fiber responsible for the NPE process. Our work differs from the original approach of Vinegoni [21] because we present the results of the PM NPE intensity dependent transmission for ultrashort (broadband) positively chirped optical pulses. Our approach allows for better understanding of how such a device will work in real all-fiber laser cavities. Selection of the fiber segmentation method significantly influences the PM NPE SA intensity dependent transmission. This behavior has been predicted by numerical simulations shown in our previous paper [28]. Here, we present detailed numerical model together with the simulated and measured power dependent transmission characteristics of the PM NPE device. The transmission of the device was measured for various input conditions, compared with the numerical modelling results and discussed.

## 2. Pulse polarization evolution in PM fibers

### 2.1 Theory and modeling

The crucial aspect of the PM NPE device design described in this paper is the choice of the scheme for segmentation and splicing of the PM fiber segments placed between two polarizing elements. Some of the possible fiber arrangements are presented in Fig. 1. When the pulse (green pulses in Fig. 1), with polarization set at a certain angle with respect to PM fiber axes, enters the device it is decomposed into two pulses with orthogonal polarizations parallel to the slow and the fast axes of the first fiber segment, respectively (green and blue in Fig. 1). The nonlinear interaction of these two pulses takes place in the presence of a strong birefringence. On a short fiber length scale, smaller than the beat length of the birefringent fiber, the birefringence will cause the linear phase shift, thus, the fiber acts like a waveplate. On the other hand, after long enough propagation the two pulses may become completely separated in time due to the GVM. In the proposed PM NPE device both: the GVM and the linear phase shift must be compensated so that only nonlinear effects contribute to the phase difference between the two pulses. The simplest way to achieve this compensation in a fiber configuration is to propagate the pulse through the exactly same length of the birefringent fiber rotated around its axis by 90° with respect to the previous one (see the second fiber segment in Fig. 1(a)) [29]. This way in each consecutive segment the two pulses exchange their role as a fast and slow one. With an even number of segments each of the pulses travels an equal distance with its polarization along the slow and fast axis and, thus, the GVM and phase shift caused by the difference of slow and fast axis refractive indices are compensated (Fig. 1(a)). Instead of splicing fiber segments with carefully matched lengths a FM can be used to achieve 90° rotation in the reflective configuration (see Fig. 1(c)). This approach was first presented for free space optics as a part of a polarization rotating compensator [30] and thereafter used in fiber setups as a part of laser cavities [18,22,25].

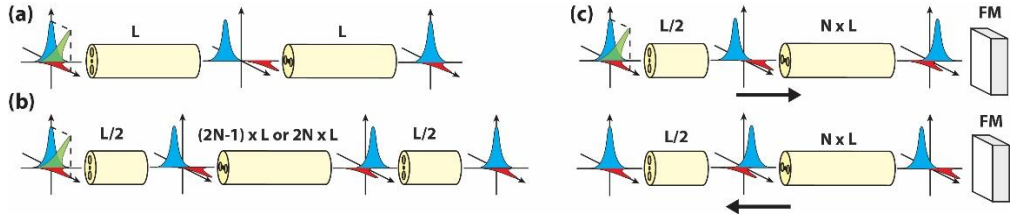

Fig. 1 Various schemes of fiber segmentation in PM NPE device: (a) two fiber segments with equal lengths, (b) odd or even number of segments (N = 1, 2, 3, …) with the first and the last segments twice shorter than the remaining ones, (c) scheme with a Faraday Mirror (FM) which is used for back reflection; upper section of Fig (c) presents the pulse propagating towards FM, lower part presents the pulse reflected by the FM. In each case the input pulse (green) can be projected on the slow (blue pulse, vertical) and fast axis (red pulse, horizontal).

In Fig. 1 different segmentation schemes for PM NPE device are presented. For the full mode-locking device the polarizer at the end of the nonlinear propagation (after the PM fiber segments) is needed (not shown in the graphics of Fig. 1). The PM NPE device serves as an artificial SA with zero transmission in the low intensity limit [21,22]. Such low small signal transmission is not the case for NOLMs with non-symmetrical coupling ratios [14,31]. Contrary to previously described all-PM-fiber cavities [22,24–26], we are not using the scheme shown in Fig. 1(a); instead we explore the advantages of using several segments of a fiber according to designs presented in Fig. 1(b-c). The major difference between the schemes presented in Fig. 1 (a) and (b) is the temporal overlap between pulses. In the configuration (a) the pulse with polarization along the slow axis (blue vertical pulse in Fig. 1(a)) is always delayed in time with respect to the other pulse (red pulse - horizontal polarization). In such a configuration the temporal overlap of the pulses is non-uniform i.e.: the leading edge of vertically polarized pulse interacts effectively only with the trailing edge of the horizontally polarized pulse and has no chance of interacting with the leading edge. This results in an asymmetrical nonlinear phase difference between the two pulses (discussed in detail in section 3.2 of the manuscript). In the configurations (b) and (c) two shorter (length L/2) fiber pieces are used at the beginning and the end of the fiber segment chain. The shorter fiber piece assures that during propagation in the next fiber segment (length L) the pulses will exchange their position in time. For the case shown in Fig. 1(b) the pulse with vertical polarization overtakes the pulse with horizontal polarization in the middle of second segment. The exchange of the position in time repeats in further optional segments with length L and finally GVM is compensated in the last L/2 fiber piece. Such an approach optimizes the time overlap of the pulses during propagation. We will show this is essential for the effective accumulation of the nonlinear phase during propagation.

The configuration (b) shows the scheme consisting of odd (even) number of segments with even (odd) number of 90° splices. The number of 90° rotations determines if the output polarization state will be parallel or orthogonal to the initial one before the angle decomposition. The scheme presented in Fig. 1(c) is realized in the reflective configuration with an FM, which corresponds to a 90° splice in a non-folded configuration. Reflective configuration always has odd number of 90° rotations, therefore, for small signal intensity it has the highest transmission for the polarization orthogonal to the input one. In this configuration the two pulses propagate through the exact same lengths of birefringent fiber segments and thus the phase shift and GVM compensation is automatically fulfilled - high accuracy of the fiber segments lengths is not required. This is not the case for the (b) configuration (also used in our previous paper [28]) in which the accuracy of the fiber segments length has to be very high (a fraction of the fiber beat length) to achieve desired transmission characteristic. The effects of various inaccuracies for the in-line configurations (Fig. 1(a-b)) were investigated numerically and are discussed in section 3.3.

Simulations were performed within the Hussar framework described in details by Kardaś et al. [32]. Briefly, the model for the NPE device is based on the Unidirectional Pulse Propagation Equation (UPPE) [33] with the use of not necessarily slowly varying envelopes for the two orthogonal polarizations:

$$E(\omega) = A_x(\omega)\hat{x}e^{i(k_x^R z - \omega_R t)} + A_y(\omega)\hat{y}e^{i(k_y^R z - \omega_R t)} \quad (1)$$

where $\omega_R$ and $k_j^R = k_j(\omega_R)$ are the reference frequency and reference values of the propagation constant, respectively ($j, l = x, y$ and $j \neq l$), $k_j(\omega)$ are the frequency dependent propagation constants with $\omega$ being the detuning from the $\omega_R$. With above definition of the electric field UPPE becomes:

$$\partial_z A_j(\omega) = i(k_j(\omega) - \frac{\omega}{v} - k_j^R)A_j(\omega) + ik_j(\omega)n_2 B_j^{NL}(\omega) \quad (2)$$

where $v$ is the simulation window velocity, and nonlinear polarization related terms are defined as:

$$\tilde{B}_j^{NL}(t) = (1 - f_R)(\tilde{B}_j^{SPM}(t) + \tilde{B}_j^{XPM}(t) + \tilde{B}_j^{DFWM}(t)) + f_R(\tilde{B}_j^{SRS}(t) + \tilde{B}_j^{XSRS}(t)) \quad (3)$$

where:

$$\tilde{B}_j^{SPM}(t) = |\tilde{A}_j(t)|^2 \tilde{A}_j(t), \tilde{B}_j^{XPM}(t) = \frac{2}{3}|\tilde{A}_l(t)|^2 \tilde{A}_j(t), \quad (4)$$

$$\tilde{B}_j^{DFWM}(t) = \frac{1}{3}\tilde{A}_l^2(t)\tilde{A}_j^*(t)e^{2i\Delta k_j z}, \tilde{B}_j^{SRS}(t) = \tilde{A}_j(t)\left((\tilde{a}(t) + \tilde{b}(t)) * |\tilde{A}_j(t)|^2\right),$$

$$\tilde{B}_j^{XSRS}(t) = \tilde{A}_j(t)\left(\tilde{a}(t) * |\tilde{A}_l(t)|^2\right) + \frac{1}{2}\tilde{A}_l(t)\left(\tilde{b}(t) * \left(\tilde{A}_j(t)\tilde{A}_l^*(t) + \tilde{A}_l(t)\tilde{A}_j^*(t)e^{2i\Delta k_j z}\right)\right)$$

are the Self-Phase Modulation (SPM), cross-Phase Modulation (XPM), Degenerate Four-Wave Mixing (DFWM; $\Delta k_j = k_l^R - k_j^R$ is the phase-mismatch), Stimulated Raman Scattering (SRS) and cross-Raman Scattering (XRS). The convolution is denoted by $(.*.)$ and $\tilde{a}(t)$ and $\tilde{b}(t)$ are the isotropic and anisotropic parts of the Raman response, respectively [34–37]. Hollenbeck et al. have given the multi-vibrational model of the Raman response function for single polarization case $(a(\omega) + b(\omega))$. We have used the frequencies and lifetimes of vibrations from this model with modified amplitudes in order to fit the experimental data [38] available for the anisotropic part $(b(\omega))$. The isotropic part $(a(\omega))$ follows from subtraction of the response function for a single polarization state and the anisotropic part. The dispersion of the fiber was characterized by $k_j^2 = 2.76 \times 10^{-2}$ ps²/m, $k_j^3 = 4.10 \times 10^{-5}$ ps³/m and $k_j^4 = -4.27 \times 10^{-8}$ ps⁴/m, where $k_j^q$ are the consecutive terms of Taylor expansion of $k_j$. The mode field diameter was 6.6 $\mu m$ and the fused silica Kerr constant $n_2 = 2.56 \times 10^{-20}$ m²/W [39] was used. The transition through the splice is in general defined by the rotation matrix. The use of envelope with reference values of the propagation constant ($k_j^R$) creates, however, the requirement for "phase update". The transition is, therefore, defined by:

$$\begin{pmatrix} A'_x \\ A'_y \end{pmatrix} = \begin{pmatrix} \cos(\theta) & \sin(\theta) \\ -\sin(\theta) & \cos(\theta) \end{pmatrix} \begin{pmatrix} A_x e^{ik_x^R L} \\ A_y e^{ik_y^R L} \end{pmatrix} \quad (5)$$

where $\theta$ is the splice angle, and the factor $e^{ik_j^R L}$ represents the linear constant phase acquired during propagation through previous segment of length L. The requirement of "phase update" comes from the definition of envelope according to Eq. 1. It has the advantage of highlighting the "phase-mismatched" processes (containing $e^{2i\Delta k_j z}$): DFWM and a part of the second term in $\tilde{B}_j^{XSRS}(t)$. However, our simulations showed that these processes have, in fact, little impact on the results (as we have verified it is below 0.2 % pulse instantaneous power and spectral energy density change upon propagation through 2 m long NPE device) and, therefore, are ignored.

The input pulse electric field in our simulations is constructed based on the measurement of spectrum and autocorrelation of the actual experimental pulse. The simulation of the propagation of a single mode (with nonlinear and linear effects considered) through fiber segments preceding and following the NPE section in an actual setup, as in the experiment, was included in order to mimic the experiment as well as possible.

## 2.2 Pulse transmission

The transmission of the NPE device can be defined twofold. First, the "total transmission" can be defined as: $T_j^T = \frac{E_j^{Out}}{E^{In}}$ (6) where $E^{In}$ and $E_j^{Out}$ ($j = x, y$) are the input and output energy of the pulse, respectively. The second definition uses "normalized peak power transmission" [22]:

$$T_j^P = \frac{P_j^{Out}}{P_x^{Out} + P_y^{Out}} \quad (7)$$

where, $P_j^{Out}$ is the peak power at the considered axis.

The major flaw of the first definition comes from the fact that the transmission defined in such way dependents heavily on the pulse shape. Both peak and wings of a Gaussian pulse can be significantly attenuated when a particular pulse is sent into the device for which a maximum transmission for a flat top pulse – with the same energy and similar duration – is expected. Moreover, this transmission depends strongly on the pulse duration. The total transmission, however, is still most useful for the comparison with experiments because usually the input pulse shape is known.

At first glance the normalized peak power transmission seems to be free of above mentioned deficiency. This is, however, not true for ultrashort pulses as the NPE device operation is then based on the interplay of the dispersion and nonlinear effects. The peak power of the pulse can decrease during propagation – due to the presence of dispersion. The nonlinear phase at the peak of a flat top pulse will also differ from that acquired by the Gaussian pulse. Therefore, the normalized peak power transmission is also pulse shape dependent. Given that and taking into account experimental limitations we have chosen to use the concept of total transmission within the present paper.

## 3. Results

*3.1 Experimental setup:*

The experimental setup for measuring the transmission of PM NPE is presented in Fig. 2. Gaussian, Fourier transform limited pulses with spectral bandwidth of 7.2 nm (centered at 1042 nm) from a femtosecond oscillator (Ybix, Lumentum) were stretched up to 4.1 ps by adding positive chirp of 0.33 ps$^2$ in a standard Martinez stretcher (not shown in Fig. 2). The laser worked at 82 MHz repetition rate. Before entering the fiber setup, the input beam propagated through a Variable Attenuator (VA) after which 5% of the beam was reflected by a Beam Splitter (BS) to constantly monitor the input power of the pulses. The positively stretched pulses were coupled by a collimator (COL$_1$) to the slow axis of a standard PM PANDA fiber (Fujikura SM98-PS-U25D-H) with $\Delta n \approx 4.5 \cdot 10^{-4}$ corresponding to the beat length of 2.4 mm and then propagated through the fiber circulator (CIR; operating exclusively in the slow axis) and Polarization Beam Splitter (PBS). The use of both CIR and PBS allowed simultaneous monitoring of the reflected light intensity in both polarizations corresponding to the slow and fast axis of the fiber, respectively. The common port of the PBS was spliced (yellow mark in Fig. 2) at a specific angle to the first PM fiber segment with the length of L/2. The angle splice defines the ratio of the intensities of the pulses in two orthogonal polarizations in the fiber. The next two fiber segments with twice the length of the first segment (L) were spliced with the 90° rotation (red mark at Fig. 2) which resulted in the change of the sign of GVM between the pulses propagating in two orthogonal axes of the PM fiber. After three fiber segments the light was coupled out by a c-lens fiber collimator (COL$_2$; 80 mm working distance) and, finally, after the double pass through the Faraday Rotator (FR) and reflection from a silver mirror (M) it was coupled back to the fiber. For small signal intensity, the setup which employs odd number of 90° splices, has zero transmission for the slow fiber axis in which the input pulse was initially launched. Increasing the peak power of the input pulse caused the change of the polarization state which results in the increased transmission for the slow axis of the fiber. Changes in the polarization state are caused entirely by the self-action of the pulse, which propagates in the birefringent fiber. Input (P$_{in}$) and output powers at the ports of CIR (slow axis) and PBS (fast axis) were monitored simultaneously. In this way the transmission of the setup could be precisely determined. Powers at both outputs of the CIR and PBS were measured with photodiode power sensors (Ophir PD300-3W). Power meters, monitoring P$_{in}$ and P$_{slow/fast\ axis}$, were cross-calibrated before each data acquisition series. The transmission values for the CIR, PBS and back reflection through the FR were measured independently to calculate the actual transmission of the PM NPE device.

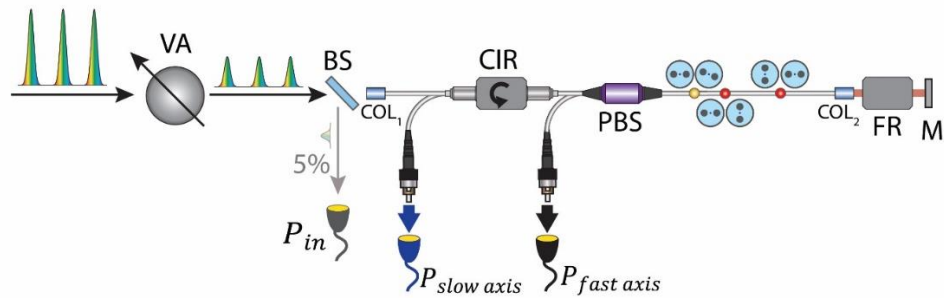

Fig. 2 The measurement setup. Positively chirped pulses with variable peak power (VA – Variable Attenuator) and time duration of 4.1 ps are divided at the input beam splitter (BS). 5% of the beam is used for continuous measurement of the input power. The remaining beam is coupled by a collimator (COL$_1$) to a PM pigtail of the polarization sensitive circulator (CIR) and propagates through Polarization Beam Splitter (PBS). After the PBS the pulses propagate through the PM NPE device; the yellow mark represents an angle splice, the two red marks represent 90° splices. The pulses are back-reflected by the silver mirror (M). A double pass

through a Faraday Rotator (FR) rotates the polarization state by 90°. The transmitted average power is measured for each polarization axis at the output of the CIR (slow axis) and the PBS (fast axis).

Fig. 3 presents the total transmission of the NPE device designed accordingly to the scheme shown in Fig. 1(c). The setup consisted of three fiber segments with the respective lengths of 1 m, 2 m and 2 m. The transmission was measured for different split ratios at the input, defined by the angle of the first splice. The first graph in Fig. 3(a) presents the simulation results for angles: 10°, 20°, 30° and 40°, which corresponds to the input pulse polarization decomposition to the slow and the fast axis of the fiber with the ratio of 0.03/0.97, 0.12/0.88, 0.25/0.75 and 0.41/0.59, respectively. Solid lines represent the transmission function of the setup for pulses propagating in the slow axis of the fiber, while dashed lines - the transmission characteristics for the fast axis. All the curves on the graph in Fig. 3(a) indicate zero transmission in the limit of very low pulse energies giving a very good contrast. The modulation depth grows and the location of the first transmission maximum shifts to higher peak powers with the increasing of the splitting ratio.

Measured transmission together with respective simulation results for the specific cases of PM NPE for 10°, 20° and 30° are presented in Fig. 3, graphs (b), (c) and (d) respectively. Simulations were performed with parameters corresponding to the experimental ones: input pulse parameters, fiber lengths and losses of the components used. The transmission was measured for the input peak power range of 0.005 – 0.68 kW. The modulation depth of the transmission curve is higher for the larger splice angle which corresponds to a more balanced intensity distribution between orthogonal polarization axes. At the same time, for smaller angle splices the first maximum in transmission is achieved for lower peak power due to a more rapid nonlinear phase difference accumulation. The full modulation depth could be achieved for the angles close to 45° where the intensities of decomposed pulses are almost the same. For this limit case, however, the location of maximum and minimum moves toward infinity along the intensity axis and thus the intensity dependence in the reasonable intensity range (e. g. bellow damage threshold) is no longer apparent. In other words: for the case of exact 45° splice, the nonlinear phase difference will be zero, thus the transmission of the setup will not be intensity dependent. The numerical simulations of the transmission are within the estimated errors of the measurements for the case of 10° and 20° input splice. For the worst case of 30° input splice, differences between measured and simulated transmissions are still less than 7.5 percentage point.

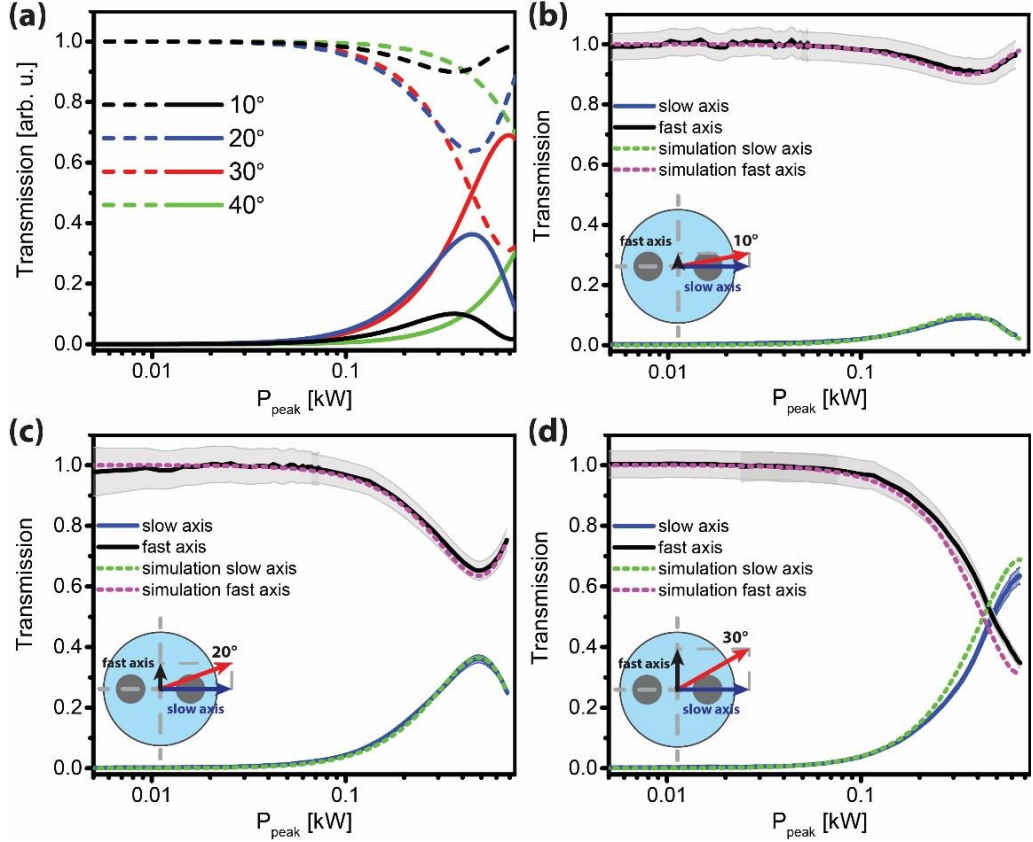

Fig. 3 The total transmission for a sample chirped Gaussian pulse (see text) and different entrance split ratios. Panel (a) presents simulations for various input splice angles. (b), (c), (d) present measured transmissions together with the simulated data for 10°, 20°, 30° input angle splices respectively. The light grey and light blue area shows the estimated absolute errors for measurements.

### 3.2 The segmentation - different fiber lengths

As we have already indicated in section 2.1, the crucial aspect in the design of the NPE PM device is the PM fiber segmentation scheme (Fig. 1). As has been shown earlier [28], a larger number of segments should result in a more symmetrical output pulse spectrum and time profile. We have simulated and measured two configurations with different number of fiber segments and the same input splice angle of 20° (split ratio 0.12/0.88). The first configuration consisted of one fiber piece only with the length of 5 m, which corresponds to the scheme represented in Fig. 4(a) which we call a non-segmented design. The second configuration was built from three fiber pieces (1 m, 2 m and 2 m) according to the scheme shown in Fig. 4(b) – further referred to as a segmented design. The yellow mark represents the input angle splice, while the red marks in the illustration represent 90° splices. The bottom graphs in Fig. 4 present the simulated phase difference accumulated by propagating pulses in two described cases.

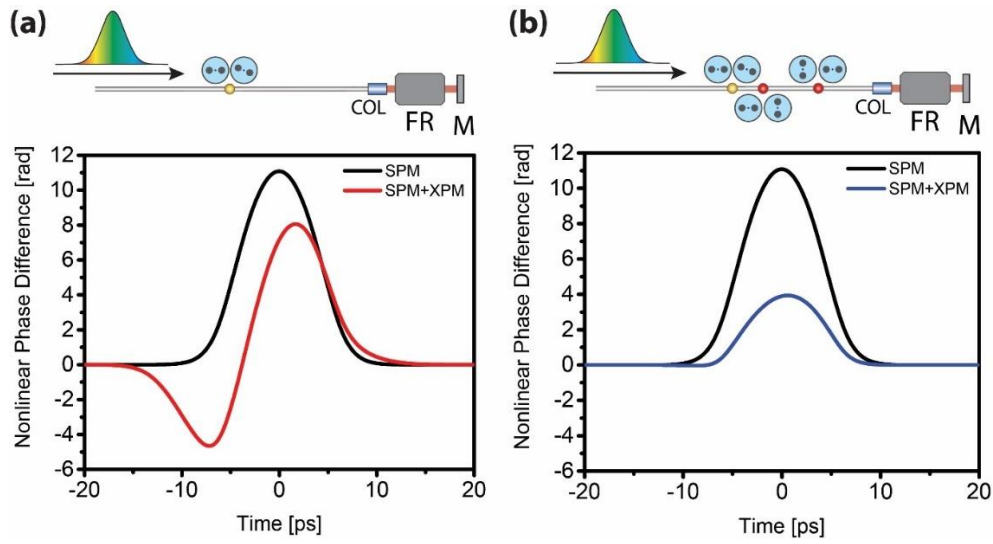

Fig. 4 Simulated nonlinear phase difference between pulses after propagation through a PM NPE device. A polarized chirped pulse propagates through the set of fiber segments than it is reflect with 90° rotation (COL – fiber collimator, FR – faraday Rotator, M – mirror). Graph (a) presents results for the non-segmented design and graph (b) shows the results for the segmented configuration. When only self-acting effects i.e.: SPM and SRS are considered the phase is an even function in time (black curves) and resemble the temporal intensity distribution of the pulse. With XPM and XSRS included, the time overlap of the pulses becomes essential. In the non-segmented design (a) an asymmetrical nonlinear phase difference represented by odd-like function in time – red curve is observed. The symmetrical (described by even-like function in time) nonlinear phase difference without crossing the zero value is achieved only for the segmented device – blue curve, graph (b).

Black curves (denoted "SPM") present the phase difference for the case in which XPM and XSRS were neglected (no interaction between pulses of orthogonal polarizations). The colored lines ("SPM+XPM") show the result obtained for the full model. In the design presented in Fig. 4(a) - red line, the phase difference between orthogonal polarizations is an odd-like function. For the second segmented device (Fig. 4(b) - blue curve), the nonlinear phase difference between orthogonal polarizations resembles the symmetrical shape of the pulse temporal profile. This is also the case for results obtained when the pulse interaction (e.g. XPM) is neglected (black curves). Therefore, the asymmetry presented in Fig. 4(a) is caused by the pulse interaction and the topology of the fiber segmentation scheme. As pointed out earlier this is explained by the decreased influence of XPM which is the main effect depending on the temporal overlap of pulses.

There is also a significant difference between the transmission characteristics of the segmented and non-segmented designs as shown in Fig. 5. For a single, long piece of a fiber, the highest transmission is achieved for lower pulse peak power of 0.2 kW with the peak transmission value of 29.4 %. On the other hand, for the PM NPE device with segmented PM fiber, higher transmission of 36.5 % was achieved for the peak power of 0.48 kW. Different properties of the two designs can be explained by the fact that the non-segmented one provides higher, although highly asymmetric, nonlinear phase difference (Fig. 5(a) and Fig. 4(a)). This is because, the role of XPM is diminished due to limited temporal overlap and, thus, SPM becomes the major contributing effect. As a result, in the non-segmented design the nonlinear phase change, which results in the pulse energy exchange between orthogonal polarizations, is achieved for the smaller input peak power. In the case of the segmented design (Fig. 5(b)) the transmission maximum is achieved for the higher peak power due to the slower nonlinear phase accumulation. The decrease of absolute value of nonlinear phase difference between orthogonal pulses is caused by the stronger influence of cross-effects which can be easily be seen by comparing of the black and blue curve in Fig. 5(b).

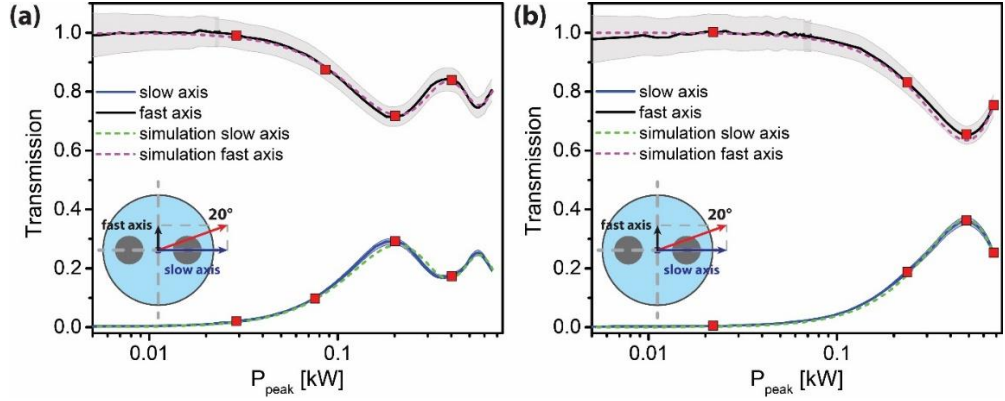

Fig. 5 Measured transmission for the non-segmented and segmented PM NPE device built accordingly to schemes presented in Fig. 4. (a) 20° - input splice and a single 5m fiber piece, (b) 20° - input splice and three fiber pieces: 1m, 2m, 2m. Red squares marks peak powers for which pulse spectra presented in Fig. 6 were measured. The light grey and light blue area shows the estimated absolute errors for measurements.

In Fig. 6 we present measured spectra (column 1), simulated spectra (column 2) and the corresponding simulated pulse temporal envelopes (column 3) after propagation through PM NPE device for both axes of the fiber. The spectra were measured with high dynamics by means of an Optical Spectrum Analyzer – Yokogawa AQ6370C. The measurements and simulations were performed for the non-segmented Fig. 6(a) and segmented device Fig. 6(b), for various input peak powers. Specific peak powers selected for those measurements are marked by red squares at the transmission measurements presented in Fig. 5. To properly describe the spectral intensity distribution of transmitted pulses we used the normalized skewness parameter – S (third statistical moment - a measure of a distribution asymmetry). Skewness was calculated only for the spectra transmitted at the slow axis. Transmission at the slow axis increases with pulse peak power which mimics SA action and can be used to start mode-locking in a fiber laser femtosecond oscillator.

The main difference between pulse parameters for the two PM NPE designs considered here can be observed for the highest transmission points (transmissions are presented in Fig. 5). For the pulse with $P_{peak} = 0.2$ kW ($P_{peak}$ - peak power) spectra for non-segmented device are highly non-symmetrical with S = 0.670 for the measured spectrum and S = 0.664 for the calculated one (Fig. 6(a) row 3). Additionally, these spectra exhibit deep valleys, which imprints severe modulations on the pulse temporal envelope. In the case of the segmented device (Fig. 6(b) row 3) we have measured and observed more symmetrical spectra with skewness lower by an order of magnitude in comparison to the results for non-segmented device. Also, the simulated pulse temporal envelope for the segmented device displays much higher quality. For the transmission maximum $P_{peak} = 0.48$ kW measured spectrum has S = 0.074 and the numerically calculated spectrum characterized with even less value of S equal 0.008. The measured spectra in both cases are in the good agreement with numerical simulations. The symmetrical nonlinear phase difference is critical for a pulse formation process, when the PM NPE device is used as a saturable absorber for broadband ultrashort optical pulses. Moreover, when such segmented PM NPE device is employed for the pulse re-shaping or optical switching, the symmetrical nonlinear phase difference is of essence.

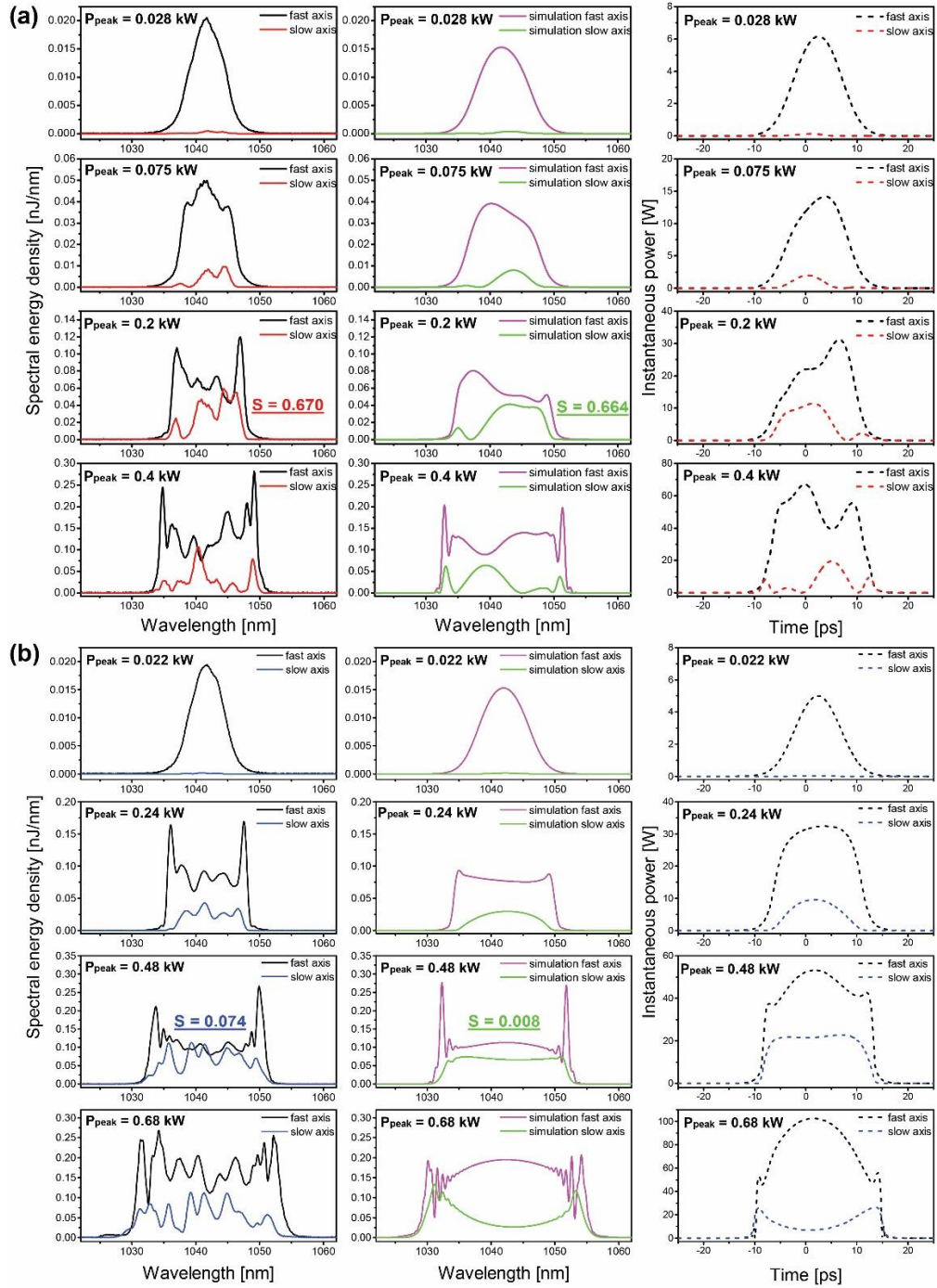

Fig. 6 Measured pulse spectra together with the simulated spectra and pulse temporal envelopes. Each row corresponds to specific pulse peak power marked by red square at the transmission characteristics presented in Fig. 5. Graph (a) presents the results for non-segmented device (Fig. 5(a)), panel (b) presents more symmetrical spectra and pulse temporal envelopes achieved for segmented device (Fig. 5(b)). S stands for normalized skewness parameter calculated for the spectra of pulses after propagation at the slow axis of PM fiber in PM NPE device.

*3.3 Phase shift - the angle and length inaccuracy*

The FM based PM NPE device automatically fulfils the condition of the GVM and phase shift compensation. Any additional birefringence induced by external conditions is also compensated. Contrary, the in-line configuration presented in Fig. 1(a-b) has more degrees of freedom. In such configuration the output angle inaccuracy as well as the fiber length inaccuracy can greatly influence the transmission of the PM NPE device. Here we present numerical simulations showing how transmission of PM NPE in-line device changes in case of input angle or segments length inaccuracies.

If the value of the second angle splice is offset by angle β in respect to the input splice, this effectively means the rotation of the output polarizer. In consequence, the peak transmission of the setup is changed. In Fig. 7(a) we present the changes in the transmission characteristic for a sample with the input angle of α=20° and 6 fiber pieces (1, 2, 2, 2, 2 and 1 m respectively) axially rotated by 90° with respect to each other. The output angle is described as: α' = α-β. By rotating the output polarizer by β = -40° we achieved very promising transmission characteristic with much higher small signal transmission and the same modulation depth. In case of β = 20° (α' = 0), however, one will observe no transmission modulation because the pulses are not projected back to the input coordinates at all. In such case there is no interference between pulses.

Potential profits of flexible adjustment of transmission through changing the output splice angle are compromised by the high fiber length accuracy requirements. In the in-line configuration the fiber lengths must be very precise within the accuracy of a small fraction of the beat length ($L_b$) of the birefringent fiber. Any difference in the length of the fiber segments can be understood as equivalent to introduction of an additional wave-plate. In graph Fig. 7(b) we present the changes of the transmission as a function of the pulse peak power for different length inaccuracies. The transmission curve is horizontally translated for higher $L_b$ errors. The error of ¼ $L_b$ corresponds to the translation of the transmission curve by ¼ of the modulation period. Similarly, the ½ $L_b$ error works as a half wave plate shifting the maximum of the transmission curve towards small intensities. The most interesting application arises for ¾ $L_b$ error (green curve). In this case the transmission curve has a positive slope in the vicinity of low peak power pulses. This feature could ease the self-starting of all-fiber ultrafast oscillators [11]. For the single- or multiple- wavelength retardation the transmission curve overlaps with the original one is not ideal (Fig. 7(b) - 1 $L_b$). This is the manifestation of the delay of the pulses caused by GVM on a distance of $L_b$.

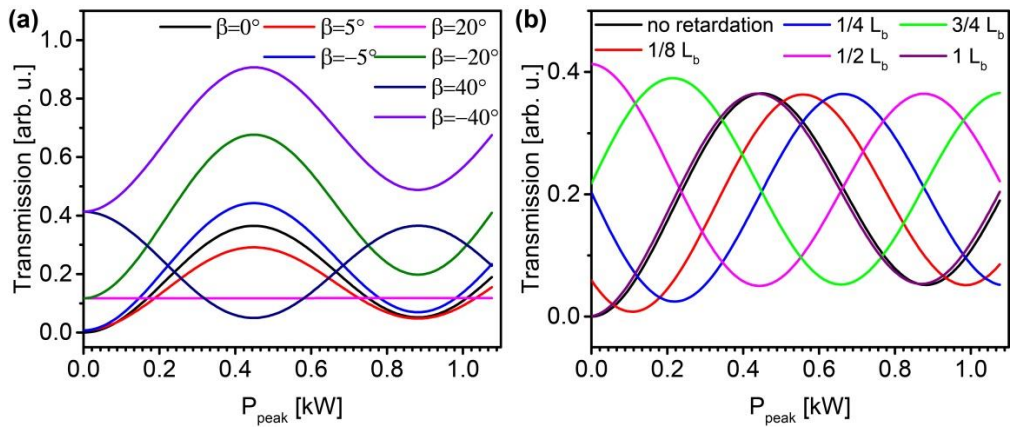

Fig. 7 Changes in transmission of the PM NPE device as a function of the output angle diversity and total fiber length inaccuracy in case of using in-line schemes without Faraday Mirror (Fig. 1(a-b)).

## 4. Conclusions

We have constructed and studied a broadband PM NPE device designed to start and maintain mode-locking in femtosecond fiber lasers. A few selected configurations were built and characterized. In addition, numerical modelling was used to study these devices. We have shown that the fiber segmentation scheme is a critical factor influencing the symmetry the output pulse spectrum. The design including shorter fiber segments at the input and output of the device (a segmented device) performs better than the one with segments of equal lengths. We have verified that the configuration based on FM alleviates the problem of high precision in the segment length critical in the in-line device. For the segmented device the transmitted spectra are characterized with an order of magnitude lower value of skewness in both measured and simulated spectra. It is apparent that the SA device which uses PM fibers, CIR and FM for back reflection is fully environmentally stable and is well suited for femtosecond fiber oscillators to be used in field. This scheme can easily be incorporated in an environmentally stable all-PM-fiber configuration by switching to the fiber coupled FM. In-line schemes, although susceptible to the inaccuracies of the fiber segment length, provide additional degrees of freedom for the SA optimization: higher peak transmission through the selection of the last splice angle as well as horizontal shifting of the transmission curve with respect to the peak power via judicious fiber length selection. The last feature can be used to optimize self-starting process in ultrafast fiber oscillators. A fiber device segmented according to the scheme proposed here can be used as an SA as well as a switching device, where the symmetrical shape of the nonlinear phase difference between two pulses must be achieved. Numerically simulated transmission characteristics of the PM NPE device agreed remarkably well with experimental results.

## Funding

The research leading to these results has received funding from National Science Centre, Poland 2016/21/N/ST2/00319 and 2017/25/B/ST7/01145 projects.

## Acknowledgments

We acknowledge P. Wasylczyk for lending the ultrafast laser source for performing the experiments.